\documentstyle{article}

 \normalmarginpar
\twocolumn
 \makeatletter
 \gdef\@proofbox{\relax}
 \long\def\proofbox#1{\gdef\@proofbox{#1}}
 \proofbox{\small{\sl PhysComp96}\\\ifx\UNDEF\@fullpaper
     Extended abstract\else Full paper\fi\\Draft, \@date}

 \gdef\fullpaper{\gdef\@fullpaper{}}

 \def\affil#1{\\{\small#1\par}}
 \gdef\@author{John Doe1\affil{No-Name University, Shipping Dept.}}
 \long\def\author#1{\gdef\@author{#1}}

 \gdef\@abstract{}
 \long\def\abstract#1{\gdef\@abstract{#1}}

\def\@maketitle{\newpage\leavevmode
  \begin{minipage}[t]{0.30\textwidth}
    \hrule height0pt
    \raggedright
    \mbox{}\par
    \@proofbox
  \end{minipage}\relax
  \begin{minipage}[t]{0.70\textwidth}
    \hrule height0pt
    \raggedleft
    \LARGE\@title\par
    \vskip4pt
    \large\@author
  \end{minipage}
  \vskip8pt
  \ifx\@abstract\@empty\else{\vskip.5em\leftskip1.5in\parskip4pt\small\@abstract
\par\vskip.5em}\fi
  \rule{\textwidth}{0.4pt}
  \vskip16pt}

\DeclareRobustCommand\em
        {\@nomath\em \ifdim \fontdimen\@ne\font >\z@
                       \upshape \else \slshape \fi}

\def\@begintheorem#1#2{\sl \trivlist \item[\hskip \labelsep{\bf #1\ #2}]}
\def\@opargbegintheorem#1#2#3{\sl \trivlist
    \item[\hskip \labelsep{\bf #1\ #2\ (#3)}]}

 \newcommand{\eq}[1]{(\ref{eq.#1})}     

 \newcommand{\sectlabel}[1]{\label{sect.#1}}
 \newcommand{\eqlabel}[1]{\label{eq.#1}}

  \def\@arabic#1{\number #1} 

\long\def\@makecaption#1#2{
        \vskip\abovecaptionskip
        \sbox\@tempboxa{{\small #1: #2}}%
        \ifdim\wd\@tempboxa>\hsize
            {\small #1: #2\par}
        \else
           \global\@minipagefalse
           \hbox to\hsize{\hfil\box\@tempboxa\hfil}
        \fi
        \vskip \belowcaptionskip}

\def\figstrut#1{\hbox to\linewidth{\vrule height#1\hfill}}

\def\comppad{\thinspace}
\def\comp{\comppad\begingroup \tt \let\do\@makeother \dospecials
          \@ifstar{\@scomp}{\@comp}}
\def\@scomp#1{\def\@tempa ##1#1{##1\endgroup\comppad}\@tempa}
\def\@comp{\obeyspaces \frenchspacing \@scomp}

\makeatother

\fullpaper

 \title{Why quantum bit commitment and \\ideal quantum coin tossing are
impossible.\thanks{Supported by DOE grant DE-FG02-90ER40542}}
 \author{Hoi-Kwong Lo\thanks{Address after 1 Oct., 96:  BRIMS,
Hewlett-Packard Labs, Filton Road, Stoke Gifford, Bristol BS12 6QZ, UK}
and H. F. Chau\thanks{Address after 1 July, 96:
Department of Physics, University of Hong Kong, Pokfulam Road,
Hong Kong}\affil{School of Natural
Sciences, Institute for Advanced Study, Princeton, NJ 08540}}
 \date{10 May, 1996\\
Revised, 21 July 1996\\
IASSNS-HEP-96/50 \\
quant-ph/9605026
}
 \abstract{

There had been well known claims of ``provably unbreakable''
quantum protocols for bit commitment and coin tossing.
However, we, and independently Mayers, showed that all proposed quantum
bit commitment (and therefore coin tossing)
schemes are, in principle, insecure because the sender,
Alice, can always cheat successfully by using an EPR-type
of attack and delaying her measurements.
One might wonder if secure quantum bit commitment and
coin tossing protocols exist at all.
Here we prove that an EPR-type of
attack by Alice will, in principle, break {\em any} realistic
quantum bit commitment and {\em ideal} coin
tossing scheme. Therefore,
provided that Alice has a quantum computer
and is capable of storing quantum signals
for an arbitrary length of time, all those
schemes are insecure.
Since bit commitment and coin tossing
are useful primitives for building up
more sophisticated protocols such as
zero-knowledge proofs, our results cast very
serious doubt on the security of quantum cryptography in
the so-called ``post-cold-war'' applications.
}

\begin{document}

\maketitle

\section{Introduction}\sectlabel{intro}

Quantum cryptography was first proposed by Wiesner \cite{Wiesner}
more than
two decades ago in a paper that remained
unpublished until 1983. Recently, there
have been lots of renewed
activities in the subject. The most well-known application of
quantum
cryptography is key distribution \cite{BB84,Bennett:SciAm,Ekert}.
The aim of key distribution is to allow two users to generate
a shared random string of information that can, for example, be
used to make their messages in subsequent communication
totally unintelligible to an eavesdropper.
Quantum key distribution is secure
\cite{Bennett:Exp,Deutsch,Lo1,Mayers1,Mayers3}
because, it is impossible (for an eavesdropper)
to make copies (or clones) of non-orthogonal states in quantum
mechanics without violating unitarity. Moreover,
measuring a quantum system generally disturbs it
because quantum mechanical observables can be non-commuting. For this
reason, eavesdropping an a quantum communication channel
will generally leave unavoidable disturbance in the transmitted
signal which can be detected by the legitimate users.

In addition to key distribution, the so-called ``post-cold-war''
applications of
quantum cryptography have also been
proposed \cite{Ar:bit,Ar:ot,BB84,BBCS92,BC91,BCJL93}.
A typical problem
in ``post-cold-war'' quantum cryptography is
the two-party secure computation, in which both parties would like
to know the result of a computation but neither side wishes to
reveal its own data. For example, two firms will embark on
a joint venture if and only if their combined capital
available for the project is larger than one million dollars. They
would like to know if this condition is fulfilled but neither
wishes to reveal the exact amount of capital it commits
to the project.
In classical cryptography, this can be done either
through trusted intermediaries or by invoking some
unproven cryptographic assumptions such as the hardness of
factoring. The big question is whether quantum cryptography
can get rid of those requirements and achieve the same goal
using the laws of physics alone.

\medskip
This paper relates to those post-cold-war applications of quantum cryptography.
Until recently, there had been much optimism in the subject.
Various protocols for say bit commitment,
coin tossing and oblivious transfer
of quantum cryptography had been
proposed \cite{Ar:bit,Ar:ot,BB84,BBCS92,BC91,BCJL93}.
In particular, the BCJL \cite{BCJL93} bit commitment scheme had been
claimed to be provably unbreakable. However, in our recent paper \cite{Lo2},
we showed that all proposed
quantum bit commitment schemes are insecure because the sender,
Alice, can always cheat successfully by using an EPR-type of
attack and delaying her measurement until she opens her commitment.
(The insecurity of the BCJL scheme was also investigated
by Mayers \cite{Mayers2} from an information-theoretic point of view.)
Our result put the security of
post-cold-war quantum cryptographic systems in serious
doubt because bit commitment is a crucial primitive
in building up more sophisticated protocols.
In particular, it has been shown by Yao \cite{Yao} that a secure quantum bit
commitment scheme can be used to implement a secure quantum
oblivious transfer scheme whereas Kilian \cite{Kilian} has shown that,
in classical cryptography, oblivious transfer can be used to
implement many protocols such as oblivious circuit evaluation,
which is
a close cousin of secure two-party computation.
This chain of arguments, therefore, seems to suggest that
quantum bit commitment alone is sufficient for implementing
secure two-party computation or its close cousin.
However, without quantum bit commitment, it is not clear if
secure two-party computation can be achieved through quantum means
at all.

\medskip
While we showed in our previous
paper \cite{Lo2} the insecurity of all proposed
quantum bit commitment schemes,
an important fundamental question
that we left unanswered was
whether {\em any} secure quantum bit commitment scheme exists at all.
Here we show that, provided that a cheater
has a quantum computer and is capable of storing quantum
signals for an arbitrary length of time,
quantum bit commitment and {\em ideal} quantum
coin tossing are impossible: All such protocols
are necessarily insecure against an EPR-type of attack by
at least one of the users.
In our opinion,
our highly disruptive results can be taken as a strong indication that,
despite widespread early optimism, realistic post-cold-war
applications of quantum cryptography simply do not exist.
We acknowledge the receipt of a
preprint of Dominic Mayers about the impossibility of quantum bit
commitment. This preprint contains the essential result and approach
to bit commitment that we present here except that, in our opinion,
it does not define in sufficient detail the general model that
it uses for quantum protocols and therefore the model is too vague.
To answer the question in a more satisfactory manner and to make
the discussion more precise, we strongly felt the need to use a variant of
the Yao's model.
Besides, our discussion on ideal quantum coin tossing
makes essential use of such a concrete model.

\section{Quantum bit commitment}\sectlabel{bit}

A general bit commitment scheme involves two parties, a sender
Alice and a receiver, Bob. Suppose
that Alice has a bit ($b = 0$ or $1$) in mind, to which she would like to
be committed towards Bob. That is to say, she wishes to provide Bob
with a piece of evidence that she has a bit in mind and that she
cannot change it. Meanwhile, Bob should not be able to tell from that
evidence what $b$ is. At a later time, however, it must be possible for Alice
to {\it open} the commitment. That is, Alice must be able to show
Bob which bit she has committed to and convinced him that this is indeed
the genuine bit that she had in mind when she
committed.\footnote{A bit commitment scheme is useful for say
implementing a coin tossing
scheme. See footnote~5 below.}
\medskip

What constitutes to a cheating by Alice? If Alice commits
to a particular value of $b$ (e.g., $b=0$) during the commitment
phase and attempts to change
it to another value (e.g., $b=1$) during the opening phase, Alice
is cheating. A bit commitment scheme is
secure against Alice only if such
a fake commitment will be discovered by Bob.
In this section, we show that, contrary to popular belief,
all quantum bit commitment schemes are, in principle, insecure against a
cheating Alice.

\subsection{Model of two-party quantum protocols}\sectlabel{two}

Quantum bit commitment and coin tossing are
examples of two-party quantum protocols.
A two-party quantum protocol involves a pair of
quantum machines
in the hands of two users, $A$ (Alice) and $B$ (Bob) respectively,
which interact with each other
through a quantum channel, $C$.
More formally, we consider
the direct product $H$ of the three Hilbert spaces $H_A$, $H_B$ and
$H_C$ where $H_A$ ($H_B$) is the Hilbert space of Alice's
(Bob's) machine and $H_C$ is the Hilbert space of the
channel. We
assume that initially each machine is in some specified pure quantum
state. $A$ and $B$ then engage in a number of rounds of quantum
communication with each other through the channel $C$. More
concretely, $A$ and $B$ alternately
performs a unitary transformation on $H_D \otimes H_C$
where $D \in \{A,B\}$.
 
\medskip

The above model
is a simplification of a model proposed
by Yao \cite{Yao}.
Although Yao apparently did not
emphasize the generality of his model,
it appears to us that any realistic two-party computation can be
described by Yao's model. For instance,
since Alice and Bob
are separated by a long distance, it is impractical
to demand simultaneous two-way communications between them.
The idea of alternate rounds of one-way communications in Yao's model
is, therefore, reasonable.
However, there are two significant distinctions
between Yao's model and ours. First, Yao's model deals
with mixed initial states whereas
we assume that the initial state of each machine is
pure. Second,
in Yao's model, the user $D$ does
two things in each round of the communication: $D$ carries
out a measurement on the current mixed state of the portion
of the space, $H_D \otimes H_C$, in his/her control and then
performs a unitary transformation on $H_D \otimes H_C$.
In our model, the measurement step has been eliminated.

\medskip

Nevertheless, we would like to argue that there is no loss
in generality and that our model still gives
the most general procedure of a two-party
quantum protocol. Let us
consider the first distinction.
In assuming that the initial state
of each machine is pure,
we are just giving the users complete control over the initial
states of the machines. Any situation with mixed initial state
can be included in our consideration simply
by attaching a quantum dice to a machine and considering
the pure state as describing the combined state of the two.

\medskip

What about the second distinction?
We make the simple but crucial observation that one can
avoid dealing with the collapse of a wavefunction
associated with a measurement altogether. The point is, that,
in principle,
$D$ has the option of adding an ancilla to his/her quantum machine and
using a {\em reversible} unitary operation to replace a measurement.
$D$ can then read off the state of his/her ancilla only
at the {\em very end} of the protocol.
Put it another way:
Alice and Bob are assumed to be in possession
of quantum computers and quantum storage devices.
Note that our model is general enough
to incorporate any classical computation and communications:
An algorithm on classical computers can clearly
be simulated by quantum computers.
It cannot be overemphasized that
this unitary description leads to no loss in
generality and indeed {\em any} two-party quantum
protocol can be described by our model. (See \cite{Griffiths}
and, in particular, the Appendix B of the revised version of \cite{Lo1}
for related discussions.)\footnote{We thank L. Goldenberg and
D. Mayers for a discussion on the generality of Yao's model.}
Such a unitary description will greatly simplify our discussion.

Of course, the faithful execution of most quantum bit commitment protocols
do not require the users to possess quantum computers.
We use a unitary description merely to simplify our discussions.
The point is the following:
If a cheater can cheat successfully against an honest
party who has a quantum computer (and quantum storage
devices), clearly he/she can also cheat
successfully against one who does not have a quantum computer
(nor quantum storage devices).
This is because
an honest party without a quantum computer can be regarded
as a special case of one who has a quantum computer but fails
to make full use of it.

\subsection{Procedure of quantum bit commitment}

Granting the possession of quantum computers and
quantum storage devices by Alice,
the most general procedure for an
ideal quantum bit commitment scheme can
be rephrased in the following manner.

\medskip

(a) Preparation of states:
Alice chooses the value of a bit $b$ to which she would like
to be committed towards Bob. If $b=0$ (respectively $b=1$), she
prepares a state
$ |0 \rangle$ (respectively $  |1 \rangle$) for $H_A$.
The two states $ |0 \rangle$ and $  |1 \rangle$
are orthogonal to each other.
Bob prepares a state $|B_0 \rangle \otimes |C_0 \rangle$ for
the product Hilbert space $H_B \otimes H_C$.
All the states $ |0 \rangle$, $  |1 \rangle$
and $|B_0 \rangle \otimes |C_0 \rangle$ are specified by the protocol
and are known to both Alice and Bob.

\medskip

(b) The actual commitment:
Step (b) involves a specified and fixed number of rounds
of quantum communication alternately between Alice and Bob.
As noted above, each
round of quantum communication can be
modeled as a unitary transformation on $H_D \otimes H_C$
($D \in \{A,B\}$), which in turn
induces a unitary transformation on the space $H= H_A \otimes H_B
\otimes H_C$. 

\medskip

Notice that for an {\em ideal} bit commitment, it must be the case
that, at the end of step (b), Bob still has absolutely no information
about the value of the committed bit $b$. (We will relax this
assumption when we come to the non-ideal case in the next subsection.)
Now that the commitment has been made, both sides may wait
an arbitrary length of time until the last step:

\medskip

(c) Opening of the commitment:
A specified and fixed number of rounds
of quantum communication alternately between
Alice and Bob are again involved.
As in step (b), we model each
round of quantum communication
as a unitary transformation on $H_D \otimes H_C$
($D \in \{A,B\}$), which in turn
induces a unitary transformation on the space $H= H_A \otimes H_B
\otimes H_C$.

\medskip

In a secure bit commitment scheme,
Bob will learn the value of $b$ and be convinced that Alice
has already committed to that value of $b$ at the end of step (b)
and cannot change it anymore in step (c).

\medskip

However, we show that the above general scheme necessarily fails because
Alice can always cheat successfully by using {\em reversible} unitary
operations in step (b) and
subsequently rotating a state that corresponds to $b=0$ to one that
corresponds to
$b=1$ and vice versa in the beginning of step (c).
Note that Alice's ability of cheating lies on her capability of
storing coherent quantum signals for a long period of
time (until the beginning of step (c)).

\medskip

Let us justify our claim. Consider more closely
the situation at the end of step (b), the commitment phase.
Let $|0 \rangle_{\rm com} $ and $|1 \rangle_{\rm com} $ denote the
state of $H = H_A \otimes H_B \otimes H_C$ at that time
corresponding to the two possible values
of $b$ respectively. In order that Alice and Bob can follow
the procedures, they must know the exact forms of all
the unitary transformations involved.\footnote{As stated
earlier, any probabilistic scheme can be rephrased as a deterministic
one by considering the state of the combined system of the quantum dice
and the original system.}
Therefore,
Alice must be capable of computing the two states
$|0 \rangle_{\rm com} $ and $|1 \rangle_{\rm com} $.
Since the channel will
sit idle for a long while, its state has to be
trivial. We may, therefore, assume
that the channel $C$ is in a prescribed pure
state $|u \rangle_C $ at the end of step (b).
Moreover, the fact that
Bob has absolutely no information about the value of $b$
implies that the density matrix in his hand is
independent of the value of $b$.
That is to say that ${\rm Tr}_{A} \left(|0 \rangle_{\rm com} 
\langle 0|_{\rm com} \right)
= {\rm Tr}_{A} \left( |1 \rangle_{\rm com} \langle 1 |_{\rm com} \right) $.
But then $|0 \rangle_{\rm com} $ and $|1 \rangle_{\rm com} $ of $H$
must have the same Schmidt polar form (See for example,
the Appendix of \cite{Hughston}.), namely:
\begin{equation}
|0 \rangle_{\rm com} = \left( \sum_k \sqrt{\lambda}_k 
| \hat{e}_k \rangle_A \otimes | \hat{\phi}_k \rangle_B \right)
 \otimes |u \rangle_C   ,
\eqlabel{polar0}
\end{equation}
and
\begin{equation}
|1 \rangle_{\rm com} = \left( \sum_k \sqrt{\lambda}_k 
| \hat{e}_k \rangle_A' \otimes | \hat{\phi}_k \rangle_B \right)
 \otimes |u \rangle_C   ,
\eqlabel{polar1}
\end{equation}
where $| \hat{e}_k \rangle_A$
and $| \hat{e}_k \rangle_A'$ are two orthonormal
bases of $H_A$ and  $ | \hat{\phi}_k \rangle_B$ 
is an orthonormal basis of $H_B$.

\medskip

The key observation is that these two states are related by
a unitary transformation acting on $H_A$ alone!
Consequently, Alice can make a fake commitment
and change the value of $b$ in the beginning of step
(c).
For example, she may proceed as follows:
First, Alice always takes $b=0$
in step (a) and goes through step (b).
It is only in the beginning of step (c) that
Alice decides on the actual value of $b$ that she wishes to
open. If she decides $b=0$ now, she can go through
step (c) honestly. If she wishes to change the
value of $b$ from $0$ to $1$, she
simply applies a unitary transformation to
rotate her state from $|0 \rangle_{\rm com}$ to $ |1 \rangle_{\rm com}$
before going through step (c).
Since the unitary transformation acts on $H_A$
alone, Bob clearly has no way of knowing Alice's
dishonesty.\footnote{What is the problem with quantum bit commitment?
Here is an analogy.
Suppose that there are two novels whose first halves are the
same, but the second halves are different. I give
you only the first half of one of the two novels and I tell you that
I have committed to a particular novel and
that I cannot change it anymore. Will you trust me?
Of course not. Since the first halves of the two novels
are the same, no real commitment has been made.
I am free to give you the second half of either novel and
claim that I have committed to either one all along. There is no way
for you to tell whether I am lying.}
In conclusion, provided that Alice possesses quantum computers
and quantum storage devices, our results show that
all quantum bit commitment schemes are insecure because
Alice can cheat successfully by using an EPR-type of attack.

\subsection{Non-ideal bit commitment}

In our above discussion, we have assumed that the bit commitment
scheme is ideal in the sense that Bob has absolutely no
information about the value of $b$ at the end of
step (b). This is the physical
reason behind the mathematical statement that
$\rho_0^{\rm com}= {\rm Tr}_{A} \left( |0 \rangle_{\rm com}
\langle 0|_{\rm com} \right)
= {\rm Tr}_{A} \left( |1 \rangle_{\rm com} \langle 1 |_{\rm com}\right) =
\rho_1^{\rm com}$.
(i.e., the two density matrices corresponding to the
two cases $b=0$ and
$b=1$ are identical.)
However, in realistic applications, one might allow Bob to have
a very tiny amount of information about $b$ at that time.
It is intuitively obvious that this is not going to
change our conclusion. On the one hand, if Bob has a large probability of
distinguishing between the two states corresponding to
$b=0$ and $b=1$ at the end of step (b), the scheme is
inherently unsafe against Bob.
On the other hand, if Bob has a small
probability of distinguishing between the two
states, then clearly, the density matrices
$\rho_0^{\rm com}= {\rm Tr}_{A} \left( |0 \rangle_{\rm com}
\langle 0|_{\rm com} \right)$ and
$\rho_1^{\rm com}= {\rm Tr}_{A} \left( |1 \rangle_{\rm com}
\langle 1 |_{\rm com} \right)$
must be
close to each other in some sense.
We have seen in the last subsection that
when the two density matrices are identical, Alice can always cheat
successfully.
It is, therefore, at least highly suggestive that, when
the two density matrices are only slightly
different, Alice will have a probability
close to $1$ of cheating successfully.
A detailed calculation, which will be sketched
briefly in the next subsection, shows that
this is indeed the case. Therefore, even non-ideal
bit commitment schemes are necessarily highly
insecure.
 
\subsection{Fidelity}

In this subsection, following
Mayers\cite{Mayers2}, we sketch the mathematical proof of the insecurity of
non-ideal quantum bit commitment scheme.
Readers who are uninterested in
mathematical details may skip this subsection on first reading.
The price that they have to pay is to take Eqs.~\eq{fidelity1} and
\eq{fidelity3} for granted.

First of all, the
{\em fidelity} \cite{Fuchs,Jozsa} between
two density matrices $\rho_0$ and $\rho_1$ of a system $B$
is defined as
\begin{equation}
F(\rho_0, \rho_1) = {\rm Tr} \sqrt {\rho_1^{1/2} \rho_0 \rho_1^{1/2}} .
\eqlabel{fidelity}
\end{equation}
$0 \leq F \leq 1$. $F=1$ if and only if $\rho_0 = \rho_1$.
Returning to the case of non-ideal bit commitment that we have
been considering,
the fact that Bob has a small
probability for distinguishing between two states $\rho_0^{\rm com}$ and
$\rho_1^{\rm com}$
implies that the fidelity $F(\rho_0^{\rm com}, \rho_1^{\rm com})$ is
very close to $1$. i.e.,
\begin{equation}
F(\rho_0^{\rm com}, \rho_1^{\rm com})= 1- \delta,
\eqlabel{largef}
\end{equation}
where $\delta$ is small.

\medskip

An alternative and equivalent
definition of fidelity involves the concept of {\em purification}.
Imagine another system $E$ attached to our given system $B$.
There are many pure states $| \psi_0 \rangle$ and $| \psi_1 \rangle$
on the composite system such that
\begin{equation}
{\rm Tr}_E \left( | \psi_0 \rangle \langle \psi_0 | \right) = \rho_0 
\mbox{~~~~~~and~~~~~}
{\rm Tr}_E \left( | \psi_1 \rangle \langle \psi_1 | \right) = \rho_1.
\eqlabel{pure}
\end{equation}
The pure states $| \psi_0 \rangle$ and $| \psi_1 \rangle$ are
called the purifications of the density matrices $\rho_0 $ and $\rho_1 $.
The fidelity can be defined as
\begin{equation}
F(\rho_0, \rho_1) = {\rm max} | \langle \psi_0|\psi_1  \rangle  |
\eqlabel{fidelity1}
\end{equation}
where the maximization is over all possible purifications.
We remark that for any fixed purification of $\rho_1$,
there exists
a maximally parallel purification of $\rho_0$ satisfying Eq.~\eq{fidelity1}.

Let us go back to a non-ideal quantum bit commitment scheme.
We take $E$ to be the combined system of Alice's machine $A$ and
the channel $C$.
It follows from Eqs.~\eq{largef} and \eq{fidelity1} that,
for the state $|1 \rangle_{\rm com}$ which is a purification of
$\rho_1^{\rm com}$,
there exist a purification $| \psi_0 \rangle_{ABC}$
of $\rho_0^{\rm com}$
such that
\begin{equation}
| \langle \psi_0| 1  \rangle_{com}  | =
F(\rho_0^{\rm com}, \rho_1^{\rm com})= 1- \delta .
\eqlabel{fidelity2}
\end{equation}

\medskip

The strategy of a cheating Alice is the same as in the ideal case.
She always prepares the state $| 0 \rangle$ corresponding to
$b=0$ in step (a) and goes through step (b).
She decides on the value of $b$ she likes only in the beginning
of the step (c). If she
chooses $b=0$, of course, she can just follow the rule.
If she chooses $b=1$, she applies a unitary transformation to
obtain the state $| \psi_0  \rangle_{ABC}$ in $H=H_A \otimes
H_B \otimes H_C$.
Notice that if she had been honest, the state would have been
$|1 \rangle_{\rm com} $ instead.
Since $| \psi_0 \rangle_{ABC}$ and $|1 \rangle_{\rm com} $
are so
similar to each other (See Eq.~\eq{fidelity2}.), Bob
clearly has a hard time in detecting
the dishonesty of Alice.
Therefore, Alice can cheat successfully with a very large probability.

\medskip

Yet another equivalent definition of the fidelity,
which will be useful in the next section, can be
given in terms of positive-operator-valued-measures (POVMs).
A POVM is a set
$\{\hat{E_b}\}$ of positive operators (i.e.,
Hermitian operators with non-negative eigenvalues) that
satisfy a sort of completeness relation
(i.e., $\sum_b \hat{E_b}$ equals the identity operator).
A POVM simply represents the most general
measurement that can be performed on a system. More concretely, it is
implemented by a) placing the system in contact with an auxiliary system
or ancilla prepared in a standard state, b) evolving the two by a
unitary operator, and c) performing an ordinary von Neumann
measurement on the ancilla.
In terms of POVMs, the fidelity is defined as
\begin{equation}
F(\rho_0, \rho_1) = {\rm min} \sum_b \sqrt{{\rm Tr} \rho_0 \hat{E_b}}
\sqrt{{\rm Tr}\rho_1 \hat{E_b}} ,
\eqlabel{fidelity3}
\end{equation}
where the minimization is over all POVMs, $\{\hat{E_b}\}$.
Eq.~\eq{fidelity3} will be useful in the next section.

\section{Quantum coin tossing}

Suppose that Alice and Bob are having a divorce and
that they are living far away from each other. They
would like to decide by a coin flip over the telephone
who is going to keep the house. Of course, if one of
them is tossing a real coin, there is no way for the other to
tell if he/she is cheating. Therefore, there must
be something else that is simulating the coin flip.
Just like bit commitment, coin tossing can be done
in classical cryptography either through trusted intermediaries
or by accepting some unproven cryptographic assumptions.
The question is:
Can quantum mechanics
help to remove those requirements? In other words,
do coin tossing schemes whose
security is based solely on the law of quantum physics
exist?

\medskip

Notice that a secure bit commitment protocol can be used trivially to
implement a secure coin tossing protocol\footnote{Alice chooses a bit
and commits it to Bob. Bob simply tells Alice
his guess for her bit. Alice then opens her commitment to
see if Bob has guessed correctly.} but the converse
is not true. Coin tossing is a weaker
protocol for which
we have a weaker result---{\em ideal\/} quantum
coin tossing schemes do not exist.
We define an ideal coin tossing scheme as one
that satisfies the following requirements:\footnote{We gratefully
thank Goldenberg and Mayers for many discussions which are
very helpful for sharpening and clarifying our ideas.}

1) At the end of the coin tossing scheme,
there are three possible outcomes: `$0$', `$1$' or `invalid'.

2) Both users know which outcome occurs.

3) If the outcomes `$0$' or `$1$' occur, Alice and Bob can be sure
that they occur with precisely the (non-zero)
probabilities, say $1/2$
each, prescribed by
the protocol.

4) If both users are honest, the outcome `invalid' will never occur.

In other words, in an {\em  ideal\/} coin tossing
scheme, both parties will always agree with each other
on the outcome. There is no room for dispute. Also,
cheating in an ideal coin tossing will only
lead to a non-vanishing probability for the occurence of
`invalid' as an outcome, but will not change the relative
probability of occurence of `0' and `1'.
Most coin tossing schemes are non-ideal.
However, any non-ideal quantum coin
tossing scheme can be regarded as an approximation
to an ideal scheme. Investigations of the ideal scheme
may, therefore, shed some lights on those more realistic,
but non-ideal, ones.
To show that ideal quantum coin tossing is impossible, we
first prove the following Lemma.

{\it Lemma}: Given that Alice and Bob initially
share no entangled quantum states, they cannot achieve ideal
quantum coin tossing without any further communication
between each other.

{\it Proof}: An ideal coin tossing scheme will give Alice
and Bob non-zero mutual information.
However, without prior classical communication, the maximal
amount of mutual information that can be gained by Alice and Bob
through local operations on shared entangled quantum states is
bounded by the entropy of formation. In the absence of
entanglement, they cannot share any mutual information.
Hence, coin tossing without prior communication
nor shared entangled quantum states must be impossible.

Now we come to the main theorem.

{\it Theorem}: Given that Alice and
Bob initially share no entangled quantum states,
{\em ideal} quantum coin tossing is impossible.

{\it Proof}: The idea of our proof is simple. We
prove by contradiction using
backward
induction. Let us assume that an ideal quantum coin tossing
can be done with a fixed and finite number, $N$, rounds of
communication between
Alice and Bob. We will prove that it can be
done in $N-1$ rounds. By repeated induction, it can
be done without any communication between Alice and Bob at all.
This is impossible because of the above Lemma.
\medskip

The induction step from $N$ rounds to $N-1$ rounds:
Suppose that there exists an ideal quantum coin tossing
protocol which involves $N$ alternate rounds of communication
between Alice and Bob. We need to prove that an ideal
quantum coin tossing protocol with only $N-1$ rounds exists.
Let us concentrate on
the $N$-th round of the communication. Without much loss
of generality, assume that it is Alice's
turn to send quantum signals through
the channel $C$ in the $N$-th round.
As this is the last
round, by the definition of
ideal coin tossing, Alice can perform a measurement and determine
the outcome, $0$, $1$ or `invalid', before
sending out the last round signals to Bob. Notice that Alice
should have no
objection against eliminating the $N$-th round altogether
because she has nothing to gain in sending the last round signals
(other than convincing Bob of the outcome).
On the other hand, Bob is supposed to learn the
outcome of the coin tossing through the combined
state in $H_B \otimes H_C$.
However, Alice, who has already known the outcome
herself, may attempt to alter Bob's outcome
by changing the mixed state in $H_C$ that she is
sending through the channel. This is essentially the
same strategy of cheating as in the case of quantum bit commitment
discussed earlier.

For the three possible outcomes in Alice's measurement,
$0$, $1$ and `invalid', let us denote
the corresponding density matrices in Bob's control
{\em before} the receipt of the $N$-th round signals
by $\rho_0^B$, $\rho_1^B$ and $\rho_{\rm invalid}^B$
respectively.
Alice's ability of cheating successfully against an honest Bob
depends on
the values of the fidelities $F(\rho_0^B, \rho_1^B)$,
$F(\rho_0^B, \rho_{\rm invalid}^B)$ and $F(\rho_1^B, \rho_{\rm invalid}^B)$. 
Here for simplicity,
we assume that there is a {\it single} pure state corresponding to
the outcome `invalid'. However, our arguments are general.
For {\em ideal\/} quantum
coin tossing, we demand the
probability of Alice cheating successfully
should be exactly zero. This implies, with the definition of
fidelity in Eq.~\eq{fidelity1}, that $F(\rho_0^B, \rho_1^B)=0$,
$F(\rho_0^B, \rho_{\rm invalid}^B)=0 $ and
$F(\rho_1^B, \rho_{\rm invalid}^B)=0 $.
It then follows from Eq.~\eq{fidelity3} that $\rho_0^B$,
$\rho_1^B$ and $ \rho_{\rm invalid}^B$
have orthogonal supports and can be completely
distinguished from one another even without the last
round of transmission from Alice. Hence,
even Bob has nothing to
gain from the last round of communication.
A truncated ideal coin tossing scheme with only
$N-1$ rounds of communication must,
therefore, be as secure
as the original $N$-round scheme.
This completes our inductive argument and we conclude that
ideal quantum coin tossing is impossible.

Unlike quantum bit commitment, for
quantum coin tossing,
there is no simple way to generalize our proof of
the impossibility of the ideal scheme the non-ideal
schemes. This is surprising because no such distinction
has been previously noted in the literature. As far as we know,
all previously proposed quantum coin tossing schemes are based
on quantum bit commitment schemes. The security of non-ideal quantum coin
tossing should be an important subject
for future investigations. We hope that our investigation
for the ideal case will shed light on the subtleties
in the non-ideal case.

\section{A constraint on two-party secure computation}

Let us consider the issue of two-party secure computation
in a more general setting.
The idea of two-party secure computation
is the following: Alice has a secret $x$ and Bob has another secret $y$.
Both would like to know the result $f(x,y)$ at the end of
a computation and be sure that the result is correct.
However, neither side wishes the other side to learn
more about its own secret than what can be deduced from the output
$f(x,y)$. As mentioned earlier, classical cryptographic
schemes can implement two-party secure computation at the cost
of introducing trusted intermediaries or accepting unproven
cryptographic assumptions.
Our results in the last two sections strongly suggest that,
in principle at least, quantum cryptography would not be
useful for getting rid of those requirements
in two-party secure computation.
Even if quantum mechanics does not help, one may ask if
there is {\em any} way
of implementing a two-party computation that is
secure from an information-theoretic point of
view? In particular, if quantum mechanics turned out to be
wrong and were replaced by a new physical theory, would it
be conceivable that two-party secure computation can
be done in this new theory?
Here we argue
that if Alice and
Bob are shameless enough to declare their
dishonesty and stop the computation whenever one of them
has a slightest advantage over the other in the amount of mutual information
he/she has on the function $f(x,y)$,
a two-party secure
computation can never be implemented.

\medskip

For simplicity, let us normalize everything and assume
that initially both Alice and Bob have no information about
$f(x,y)$ and at the end of the computation, both have
$1$ bit of information about $f(x,y)$. Let us suppose
further than Alice and Bob are unkind enough to
stop the computation whenever one of them has an $\epsilon$ bit
of information more than the other.
Any realistic scheme must involve a finite number say $N$ alternate
rounds of communication between Alice and Bob.
An analogy is that two persons, Alice
and Bob, are walking in discrete alternate steps from the starting
point $0$ to the finishing line set at $1$.
Altogether $N$ steps are made and it is demanded that
Alice and Bob will never be separated from each other
for more than a distance $\epsilon$.\footnote{Actually,
there is a minor subtlety in quantum cryptography.
Each time when one user, say Alice, advances,
the other user, say Bob, may slip backwards. The point is:
the quantum ``no-cloning theorem'' states
that quantum signals cannot be copied. When Bob sends signals
to Alice, he loses control over the signals that he sends.
In other words, Bob's available information
tends to decrease whenever he sends signals to Alice.}
Clearly, this implies $N \epsilon \geq 1$ or $N \geq 1/\epsilon$.
Therefore, the smaller the tolerable
relative informational advantage $\epsilon$ is, the larger the number of
rounds of communication $N$ is needed.
Notice that the constraint $N \epsilon \geq 1$ applies
to {\em any} two-party secure computation scheme. In particular,
it remains valid even if quantum mechanics is wrong.

It may also be of some interest to speculate that a similar inequality
$N \epsilon \geq 0(1) $ may hold for non-ideal quantum coin tossing schemes
where $N$ is the number of rounds of communication and $\epsilon$ is
the probability that a user cheats successfully. Consequently, as
$\epsilon \to 0$, $N \to \infty$ and ideal quantum coin tossing
with finite rounds of communication becomes impossible.

\section{Summary}

We have shown that all realistic
quantum bit commitment and ideal quantum coin tossing protocols
are, {\em in principle}, insecure.
The basic problem is that the users
can cheat using an EPR-type of attack.
Our results totally contradict well-known
claims of ``provably unbreakable'' schemes
in the literature, whose analyses on EPR attack
were flawed, and provide strong evidence
against the
security of quantum cryptography in ``post-cold-war''
applications, at least
{\em in principle}. The early optimism in the subject is, therefore,
misplaced. Nevertheless, quantum bit commitment schemes that are
secure {\em in practice}
may still exist\footnote{We thank C. H. Bennett for a
discussion about this point.} because it is notoriously difficult for
cheaters with current technology to store quantum signals for an arbitrary
length of time. A more serious consideration is
the following. In order to cheat successfully,
a cheater may need a quantum computer, but
such a computer is not yet available with current technology.
Therefore, we can trade the traditional complexity assumption
with an assumption on the inability of the cheater to store quantum
signals for a long period of time and to
build and operate a quantum computer. This subject deserves further
investigations.
Another important unsolved problem is
the security of non-ideal quantum coin tossing.
Finally, we remark that,
thanks to the quantum ``no-cloning'' theorem,
the security of quantum
{\em key distribution}\cite{Deutsch,Lo1,Mayers1,Mayers3}
is widely accepted and
quantum cryptography is useful at least for
this purpose. We expect that quantum key distribution will remain
a fertile subject for years to come.

\section{Acknowledgments}

Numerous helpful discussions with M. Ardehali, C. A. Fuchs,
J. Kilian, J. Preskill, M. H. Rubin, L. Vaidman,
A. Wigderson, F. Wilczek, and A. Yao are
acknowledged. We are very grateful to L. Goldenberg and D. Mayers
for pointing out a crucial error in the
discussion of quantum coin tossing
in an earlier version of this manuscript and
for many discussions which are very helpful
for sharpening and clarifying our ideas.
We also acknowledge receipt of
preprints
on the impossibility of quantum bit commitment from D. Mayers
and on coin tossing from L. Goldenberg.
We also thank C. H. Bennett, D. Divincenzo, R. Jozsa and T. Toffoli
for generous advice and useful comments.

\end{document}